\documentclass[a4]{article}
\usepackage[T1]{fontenc}
\usepackage[latin1]{inputenc}
\usepackage{amsmath, amssymb, amsthm}
\usepackage[english]{babel}
\usepackage{indentfirst}
\usepackage{booktabs}
\usepackage{array}
\usepackage{caption,appendix}
\usepackage{geometry}
\usepackage{float}
\usepackage{cancel}
\usepackage{bbold}
\usepackage{authblk}
\usepackage{cite}
\usepackage{bm}
\usepackage{xcolor}
\allowdisplaybreaks
\numberwithin{equation}{section}

\begin{document}
\author[1,2,3]{Salvatore Capozziello \thanks{capozziello@na.infn.it}}
\author[4]{Maurizio Capriolo \thanks{mcapriolo@unisa.it} }
%\author[4,5]{Gaetano Lambiase \thanks{glambiase@unisa.it}}
\affil[1]{\emph{Dipartimento di Fisica "E. Pancini", Universit\`a di Napoli {}``Federico II'', Compl. Univ. di
		   Monte S. Angelo, Edificio G, Via Cinthia, I-80126, Napoli, Italy, }}
\affil[2]{\emph{INFN Sezione  di Napoli, Compl. Univ. di
		   Monte S. Angelo, Edificio G, Via Cinthia, I-80126,  Napoli, Italy,}}
\affil[3]{\emph{Scuola Superiore Meridionale, Largo S. Marcellino 10,  I-80138,  Napoli, Italy,}}
\affil[4]{\emph{Dipartimento di Fisica Universit\`a di Salerno, via Giovanni Paolo II, 132, Fisciano, SA I-84084, Italy.} }
%\affil[5]{\emph{INFN Sezione  di Napoli,Gruppo Collegato di Salerno, via Giovanni Paolo II, 132, Fisciano, SA I-84084, Italy.} }
\date{\today}
\title{\textbf{Gravitational waves in $f(Q)$ non-metric gravity without gauge fixing }}
% $f\left(R,\Box^{-1}R\right)$ gravity}}
\maketitle
\begin{abstract} 
We investigate the polarization modes of gravitational waves  in $f(Q)$ non-metric gravity without gauge fixing.  The main result of this study is that no further scalar mode appears more than  the two standard plus and cross transverse polarizations of  massless tensor gravitational radiation, typical of General Relativity. This is because the first-order perturbation of  connection does not modify the linearized field equations in vacuum which remain gauge invariant.  Then,  the world line equations  of free point particles, as well as the equations of their deviations, are obtained using only the symmetric teleparallel connection.   In $f(Q)$ gravity, test masses follow timelike geodesics and not autoparallel curves. In the proper reference frame, thanks to the geodesic deviation equation  of the structure-less bodies in free fall, we  prove that, in any gauge,  only the metric perturbations $h_{\mu\nu}$, related to   tensor modes, survive by exploiting the gauge invariance. Besides, scalar modes   disappear.  This allows us to conclude that only two degrees of freedom of linearized $f(Q)$ non-metric gravity propagate as in General Relativity and  in $f(T)$ teleparallel gravity.  The situation is different with respect to $f(R)$ gravity (with $f(R)\neq R$) where a further scalar mode is found.
\end{abstract}
\section{Introduction}
General Relativity (GR) is the theory describing gravitational interaction according to the geometric description formulated by Einstein and physically based on the validity of the Equivalence Principle. 
Confirmations of this theory range from Solar System dynamics, to the existence of gravitational waves and black holes up to the dynamical formulation of the cosmological problem. However,   despite the undeniable successes of this theory, several shortcomings emerge at infrared and ultraviolet scales.  To overcome these issues, extensions and modification of  the Einstein picture  have been developed. See, e.g. ~\cite{Starobinsky, Nojiri1, Caprep, Nojiri2, Clifton, DeFelice, Faraoni, Odi2,Cai}.  Some of these are  based on metric-affine geometry where the spacetime manifold is endowed with a metric tensor and  a general connection,  which can be  also not compatible with metric, and features both torsion and curvature \cite{Carmen1}.  These two geometric objects, metric and connection,  are treated as independent, i.e., as fields whose evolution describes gravitational interaction via equations of motion  arising from the given theory  of gravity.  

In principle, the number of degrees of freedom of metric-affine theories is of $10$ components for the metric $g_{\mu\nu}$ and $64$ for the affine connection $\Gamma^{\alpha}_{\phantom{\alpha}\mu\nu}$, necessary to encode the gravitational field.  If we limit ourselves to connections without torsion and  curvature, as in  symmetric teleparallel  gravity  (STG) theories, this number is drastically reduced to $10+4$, for details see Ref.~\cite{Lavinia}.  A straightforward   extension of symmetric teleparallel equivalent General Relativity (STEGR), described by  a linear action of the non-metricity scalar $Q$,  there is the so called  $f(Q)$ gravity, where $f$ is a generic analytic function.  As we will demonstrate in this study,   only two degrees of freedom, responsible of  two tensor modes, propagate without fixing any gauge. In \cite{Taishi},  the scalar mode in $f(Q)$ gravity appears as a ghost, and can be eliminated by a constraint, and therefore does not propagate. However, the number of degrees of freedom in STG theories, and, in particular,  the number of propagating modes, i.e. the physical ones that cannot be eliminated, is matter of debate and it is still an open question. In Ref.~\cite{Hu:2022anq}, authors claim that the degrees of freedom in $f(Q)$ gravity are 8, while, in Ref.~\cite{DAmbrosio:2023asf},  it is argued that the degrees of freedom should be equal to or less than 7. 

In this paper, we investigate, in the framework of $f(Q)$ non-metric gravity, the possible presence of further  modes in  gravitational radiation beyond the two standard tensor ones of GR.  We will perform the analysis in   arbitrary gauge, without fixing  the coincident gauge as done in many studies~\cite{Sebastian,Jose,XLHL,NF,CONKOI,BJHK}. 

We  will first derive the  world-line equations of  test bodies and then the  equation of deviations. After,  we discuss the GW polarization modes by metric perturbations $h_{\mu\nu}$ expressed in any gauge.  Some astrophysical and cosmological applications of $f(Q)$ gravity are reported in~\cite{App1,App2,App3,App4,App5,App6,App7, NoOdi}.  A different approach to  GWs is reported in~\cite{HPUS, SFSGS}. In order to study  the production of GWs,  it is necessary to derive the related gravitational energy-momentum pseudo-tensor for.    This object is derived  in \cite{CCL1} for non-local theories of gravity.  Pseudo-tensors for higher-order curvature-based gravity are discussed in \cite{CCL, CCT, ACCA}, while, for   teleparallel  gravity, see~\cite{Capozziello:2018qcp}.  The propagation of GWs in non-local gravity is developed in Refs.~\cite{CAPRIOLOM, CCCQG2021, CCN}. GWs  in teleparallel  and in metric  gravity are discussed in~\cite{CCC, CCC1, CCC2}. 

The present paper has the following structure.  Sec.~\ref{A} is devoted to a summary of the main geometric objects  characterizing  non-metric gravity. In particular, we  underline  the difference between autoparallel and geodesic curves. After deriving the field  equations  from a variational principle in Sec.~\ref{B}, we will  linearize them  in any gauge in  Sec.~\ref{C}. Specifically, we use   the linear perturbations of metric tensor and STG connection.  Thanks to the wave solutions,  derived in Sec.~\ref{D} via  the Fourier transformations,  the geodesic deviation equation can expressed in a local Lorentz frame (see Sec.~\ref{E}).  It can be solved in the metric perturbation $h_{\mu\nu}$ expressed in an arbitrary gauge. See Sec.~\ref{F}.  Finally the polarizations and the nature of GW modes is discussed in Sec.~\ref{F}. Discussion and  conclusions are reported in Sec.~\ref{G}.  The fall of free body  and geodesic deviation equations are derived in the two final appendices.

Adopted conventions  are:  $\eta_{\mu\nu}=\text{diag}(1,-1,-1,-1)$ for the Minkowski metric tensor;   $R^{\mu}_{\phantom{\mu}\nu\alpha\beta}=\partial_{\alpha}\Gamma^{\mu}_{\phantom{\mu}\nu\beta}\dots$ for the Riemann tensor;  $R_{\nu\beta}=R^{\lambda}_{\phantom{\lambda}\nu\lambda\beta}$ for the Ricci tensor.  Along the draft,  $\Gamma$ stands for the teleparallel symmetric connection as well as $\nabla$ stands for covariant derivative with respect to the non-metric connection $\Gamma$.

\section{A geometric summary}\label{A}
In an arbitrary metric-affine geometry, we can uniquely decompose the affine connection $\Gamma^{\alpha}_{\phantom{\alpha}\mu\nu}$ in  presence of curvature, torsion and non-metricity, into three parts as~\cite{RV,JHK, DKC, CDF} 
\begin{equation}\label{1}
\boxed{
\Gamma^{\alpha}_{\phantom{\alpha}\mu\nu}=\genfrac\{\}{0pt}{1}{\alpha}{\mu\nu}+K^{\alpha}_{\phantom{\alpha}\mu\nu}+L^{\alpha}_{\phantom{\alpha}\mu\nu}
}\ ,
\end{equation}
where $\genfrac\{\}{0pt}{1}{\alpha}{\beta\gamma}$ are the \emph{Christoffel symbols} defined  as
\begin{equation}\label{1_1}
\boxed{
\genfrac\{\}{0pt}{1}{\rho}{\mu\nu}\equiv\frac{1}{2}g^{\rho\alpha}\left(\partial_{\nu}g_{\mu\alpha}+\partial_{\mu}g_{\nu\alpha}-\partial_{\alpha}g_{\mu\nu}\right)
}\ ,
\end{equation}
$K^{\alpha}_{\phantom{\alpha}\mu\nu}$ is the {\em contortion tensor} defined as 
\begin{equation}\label{4}
\boxed{
K^{\alpha}_{\phantom{\alpha}\mu\nu}:=\frac{1}{2}g^{\alpha\lambda}\left(T_{\mu\lambda\nu}+T_{\nu\lambda\mu}+T_{\lambda\mu\nu}\right)}\ ,
\end{equation}
through the {\em torsion tensor} $T^{\alpha}_{\phantom{\alpha}\mu\nu}$
\begin{equation}\label{3}
T^{\alpha}_{\phantom{\alpha}\mu\nu}=2\Gamma^{\alpha}_{\phantom{\alpha}[\mu\nu]}\ ,
\end{equation}
and $L^{\alpha}_{\phantom{\alpha}\mu\nu}$ is the {\em disformation tensor} defined as 
\begin{equation}\label{7}
\boxed{
\begin{aligned}
L^{\alpha}_{\phantom{\alpha}\mu\nu}&:=-\frac{1}{2}g^{\alpha\lambda}\left(Q_{\mu\lambda\nu}+Q_{\nu\lambda\mu}-Q_{\lambda\mu\nu}\right)\\
&=\frac{1}{2}Q^{\alpha}_{\phantom{\alpha}\mu\nu}-Q_{(\mu\phantom{\alpha}\nu)}^{\phantom{(\mu}\alpha}
\end{aligned}
}\ ,
\end{equation}
by means of $Q_{\alpha\mu\nu}$ which is the {\em non-metricity tensor} defined as 
\begin{equation}\label{9}
\boxed{
Q_{\alpha\mu\nu}=\nabla_{\alpha}g_{\mu\nu}=\partial_{\alpha}g_{\mu\nu}-\Gamma^{\beta}_{\phantom{\beta}\alpha\mu}g_{\beta\nu}-\Gamma^{\beta}_{\phantom{\beta}\alpha\nu}g_{\beta\mu}
}\ .
\end{equation}
The contortion tensor is antisymmetric in the first and third indices 
\begin{equation}\label{5}
K_{\alpha\mu\nu}=-K_{\nu\mu\alpha}\ ,
\end{equation}
and its antisymmetric part is given by
\begin{equation}\label{6}
K^{\alpha}_{\phantom{\alpha}[\mu\nu]}=\frac{1}{2}T^{\alpha}_{\phantom{\alpha}\mu\nu}\ .
\end{equation}
The disformation tensor $L^{\alpha}_{\phantom{\alpha}\mu\nu}$ is symmetric in the second and third indices 
\begin{equation}\label{8}
L^{\alpha}_{\phantom{\alpha}[\mu\nu]}=0\ ,
\end{equation}
and the non-metricity tensor $Q_{\alpha\mu\nu}$ is symmetric in the last two indices  
\begin{equation}\label{10}
Q_{\alpha[\mu\nu]}=0\ ,
\end{equation}
and furthermore satisfies the following Bianchi identity
\begin{equation}\label{8.8}
\nabla_{[\alpha}Q_{\beta]\mu\nu}=0\ .
\end{equation}
 {\em Symmetric teleparallel gravity} (STG) represents   a class of non-metric theories of gravity where both the curvature and the torsion of the affine connection vanish, and therefore it is a formulation of gravity described  in terms of non-metricity only, i.e. ,
\begin{equation}\label{11}
\boxed{0=R^{\lambda}_{\phantom{\lambda}\mu\nu\sigma}}=\Gamma^{\lambda}_{\phantom{\lambda}\mu\sigma,\nu}-\Gamma^{\lambda}_{\phantom{\lambda}\mu\nu,\sigma}+\Gamma^{\beta}_{\phantom{\beta}\mu\sigma}\Gamma^{\lambda}_{\phantom{\lambda}\beta\nu}-\Gamma^{\beta}_{\phantom{\beta}\mu\nu}\Gamma^{\lambda}_{\phantom{\lambda}\beta\sigma}\ ,
\end{equation}
and
\begin{equation}\label{12}
\boxed{
T^{\alpha}_{\phantom{\alpha}\mu\nu}=0
}\ .
\end{equation}
Therefore, in STG, the connection reads as
\begin{equation}\label{13_0}
\boxed{
\Gamma^{\alpha}_{\phantom{\alpha}\mu\nu}=\genfrac\{\}{0pt}{1}{\alpha}{\mu\nu}+L^{\alpha}_{\phantom{\alpha}\mu\nu}
}\ .
\end{equation}
The {\em non-metricity scalar} is
\begin{equation}\label{14}
\boxed{
\begin{aligned}
Q\equiv&-\frac{1}{4}Q_{\alpha\beta\gamma}Q^{\alpha\beta\gamma}+\frac{1}{2}Q_{\alpha\beta\gamma}Q^{\gamma\beta\alpha}+\frac{1}{4}Q_{\alpha}Q^{\alpha}-\frac{1}{2}Q_{\alpha}\widetilde{Q}^{\alpha}\\
=&\, g^{\mu\nu}\Bigl(L^{\alpha}_{\phantom{\alpha}\beta\mu}L^{\beta}_{\phantom{\beta}\nu\alpha}-L^{\alpha}_{\phantom{\alpha}\beta\alpha}L^{\beta}_{\phantom{\beta}\mu\nu}\Bigr)
\end{aligned}
}\ ,
\end{equation}
and its two  traces are
\begin{equation}\label{15}
Q_{\alpha}\equiv Q_{\alpha\phantom{\mu}\mu}^{\phantom{\alpha}\mu}\ ,
\end{equation}
and
\begin{equation}\label{16}
\widetilde{Q}^{\alpha}\equiv Q_{\mu}^{\phantom{\mu}\mu\alpha}\ .
\end{equation}
Accordingly,  the contractions of the disformation tensor $L^{\lambda}_{\phantom{\lambda}\alpha\beta}$ become
\begin{equation}
L^{\lambda}_{\phantom{\lambda}\alpha\lambda}=-\frac{1}{2}Q_{\alpha}\quad\ ,\quad
L^{\alpha\lambda}_{\phantom{\alpha\lambda}\lambda}=\frac{1}{2}Q^{\alpha}-\widetilde{Q}^{\alpha}\ .
\end{equation}
We  can introduce a  useful geometric object, the {\em non-metricity conjugate tensor}, defined as 
\begin{equation}\label{17}
\boxed{
P^{\alpha}_{\phantom{\alpha}\mu\nu}=\frac{1}{2}\frac{\partial Q}{\partial Q_{\alpha}^{\phantom{\alpha}\mu\nu}}
}\ ,
\end{equation}
that, thanks to the symmetry of $Q_{\alpha\mu\nu}$~\eqref{10}, is symmetric in the last two indices
\begin{equation}
P^{\alpha}_{\phantom{\alpha}\mu\nu}=P^{\alpha}_{\phantom{\alpha}(\mu\nu)}\ .
\end{equation}
The superpotential $P^{\alpha}_{\phantom{\alpha}\mu\nu}$ defined in Eq.~\eqref{17}, thanks to  the non-metricity scalar $Q$ given in~\eqref{14}, can be written as 
\begin{eqnarray}\label{24}
P^{\alpha}_{\phantom{\alpha}\mu\nu}&=&\frac{1}{4}\Bigl[-Q^{\alpha}_{\phantom{\alpha}\mu\nu}+2Q_{(\mu\phantom{\alpha}\nu)}^{\phantom{(\mu}\alpha}+Q^{\alpha}g_{\mu\nu}-\widetilde{Q}^{\alpha}g_{\mu\nu}-\delta^{\alpha}_{(\mu}Q_{\nu)}\Bigr]\nonumber\\
&=&-\frac{1}{2}L^{\alpha}_{\phantom{\alpha}\mu\nu}+\frac{1}{4}\bigl(Q^{\alpha}-\widetilde{Q}^{\alpha}\big)g_{\mu\nu}-\frac{1}{4}\delta^{\alpha}_{(\mu}Q_{\nu)}\\
&=&-\frac{1}{2}L^{\alpha}_{\phantom{\alpha}\mu\nu}+\frac{1}{4}B^{\alpha}g_{\mu\nu}-\frac{1}{4}\delta^{\alpha}_{(\mu}Q_{\nu)}\ ,
\end{eqnarray}
where the vector $B^{\alpha}$ is defined as 
\begin{equation}
B^{\alpha}=Q^{\alpha}-\widetilde{Q}^{\alpha}\ .
\end{equation} 
This allows us to rewrite the non-metricity scalar Q as
\begin{equation}\label{18}
Q=Q_{\alpha\mu\nu}P^{\alpha\mu\nu}\ .
\end{equation}
\subsection{Autoparallel curves and geodesics}
The {\em autoparallel curves} or affine geodesics are the {\em straightest  curves}, that is,  a curve $x^{\mu}(s)$ along which the tangent vector $t=t^{\mu}{\bf e}_{\mu}=\bigl(dx^{\mu}/ds\bigr){\bf e}_{\mu}$ is parallel transported with respect to the connection $\Gamma$,  namely if
\begin{equation}\label{18_1}
\nabla_{t}t=0\longleftrightarrow t^{\alpha}\nabla_{\alpha}t^{\mu}=0\longleftrightarrow \boxed{\frac{d t^{\alpha}}{ds}+\Gamma^{\alpha}_{\phantom{\alpha}\beta\gamma}t^{\beta}t^{\gamma}=0}\ .
\end{equation}
Then, the autoparallel curves only require the connection on the manifold.
The {\em geodesic} or metric geodesic are the {\em "shortest"  curves} connecting two given points,  that is,  let $s$ be the distance along the timelike curve $x^{\mu}(s)$, the geodesic is the local extreme of the length functional $s$ obtained through the Euler-Lagrange equations, namely 
\begin{equation}\label{18_2}
\delta s=\delta\int_{s_{1}}^{s_{2}} \sqrt{g_{\mu\nu}\frac{dx^{\mu}}{ds}\frac{dx^{\nu}}{ds}}ds=0\longleftrightarrow \boxed{\frac{d t^{\alpha}}{ds}+\genfrac\{\}{0pt}{1}{\alpha}{\beta\gamma}t^{\beta}t^{\gamma}=0}\ ,
\end{equation}
where $\genfrac\{\}{0pt}{1}{\alpha}{\beta\gamma}$ are the  Christoffel symbols defined in Eq.~\eqref{1_1}. Thus  Eq.~\eqref{18_2} is the so-called {\em geodesic equation}.
Then the geodesic curves  require only the  metric tensor  defined on the manifold. 
In GR,  the connection is metric-compatible and symmetric. From Eq.~\eqref{1}, we have 
\begin{equation}\label{18_3}
\Gamma^{\alpha}_{\phantom{\alpha}\beta\gamma}=\genfrac\{\}{0pt}{1}{\alpha}{\beta\gamma}\rightarrow\Gamma^{\alpha}_{\phantom{\alpha}\beta\gamma}=\hat{\Gamma}^{\alpha}_{\phantom{\alpha}\beta\gamma}\ ,
\end{equation}
that is the connection $\Gamma^{\alpha}_{\phantom{\alpha}\beta\gamma}$ becomes the {\em Levi-Civita connection} $\hat{\Gamma}^{\alpha}_{\phantom{\alpha}\mu\nu}$, namely 
\begin{equation}\label{2}
\boxed{
\hat{\Gamma}^{\alpha}_{\phantom{\alpha}\mu\nu}:=\frac{1}{2}g^{\alpha\lambda}\left(\partial_{\mu}g_{\lambda\nu}+\partial_{\nu}g_{\lambda\mu}-\partial_{\lambda}g_{\mu\nu}\right)
}\ .
\end{equation}
Therefore  autoparallel curves and  geodesics exactly coincide only in GR. For a generic non-metric connection,  this statement does not work. As we will see, in STGs as $f(Q)$ gravity,  autoparallel curves and  geodesics do not coincide. This means that  falling bodies follow  geodesics and not  autoparallel curves. 

\section{Field and connection equations}\label{B}
In STG, spacetime is a Lorentz manifold equipped with a metric tensor $g_{\mu\nu}$ and a non-metric, torsion-free and curvature-free connection, the symmetric teleparallel connection $\Gamma^{\alpha}_{\phantom{\alpha}\mu\nu}$ as defined in~\eqref{13_0}. Metric and connection are   governed by the respective equations of motion.  From now on, we will use only the non-metric, symmetric and teleparallel connection without ever using the Levi-Civita one. 

The action in  $f(Q)$  gravity is expressed in the framework of Palatini formalism.  Here $g_{\mu\nu}$  and $\Gamma^{\alpha}_{\phantom{\alpha}\mu\nu}$ are independent dynamical  variables \cite{JHKP,DZ}.  It is 
\begin{equation}\label{13}
S_{f(Q)}=\int_{\Omega}d^{4}x\,\Bigl[\frac{1}{2\kappa^{2}}\sqrt{-g}f\left(Q\right)+\lambda_{\alpha}^{\phantom{\alpha}\beta\mu\nu}R^{\alpha}_{\phantom{\alpha}\beta\mu\nu}+\lambda_{\alpha}^{\phantom{\alpha}\mu\nu}T^{\alpha}_{\phantom{\alpha}\mu\nu}+\sqrt{-g}\mathcal{L}_{m}\bigl(g\bigr)\Bigr]\ ,
\end{equation}
where $\kappa^{2}=8\pi G/c^{4}$ and $\lambda_{\alpha}^{\phantom{\alpha}\beta\mu\nu}=\lambda_{\alpha}^{\phantom{\alpha}\beta[\mu\nu]}$, $\lambda_{\alpha}^{\phantom{\alpha}\mu\nu}=\lambda_{\alpha}^{\phantom{\alpha}[\mu\nu]}$ are  Lagrange multipliers.
% i.e., other 96 plus 24 independent scalar fields. 
Varying  the action $S_{f(Q)}$ with respect to  the metric tensor $g_{\mu\nu}$, we have~\cite{DKC, JHKP,DZ, CDA, BVC, CFSM, VCCE,   Esposito, ABS, Mehdi}
	\begin{equation*}
	\delta_{g}S_{f(Q)}=0\ .
	\end{equation*}
By requiring that the variation of  metric tensor vanishes on the domain boundary  $\Omega$, we obtain the  the   field equations:
\begin{equation}\label{22}
\boxed{
\frac{2}{\sqrt{-g}}\nabla_{\alpha}\left(\sqrt{-g}f_{Q}P^{\alpha}_{\phantom{\alpha}\mu\nu}\right)-\frac{1}{2}g_{\mu\nu}f+f_{Q}\Bigl(P_{\mu\alpha\beta}Q_{\nu}^{\phantom{\nu}\alpha\beta}-2Q^{\alpha\beta}_{\phantom{\alpha\beta}\mu}P_{\alpha\beta\nu}\Bigr)=\kappa^{2}T_{\mu\nu}
}\ ,
\end{equation} 
where $f_{Q}=\partial f/\partial Q$ and $T_{\mu\nu}$ is the matter energy-momentum tensor  defined as
\begin{equation}
T_{\mu\nu}=-\frac{2}{\sqrt{-g}}\frac{\delta \bigl(\sqrt{-g}\mathcal{L}_{m}(g)\bigr)}{\delta g^{\mu\nu}}\ .
\end{equation} 
Varying the action~\eqref{13} with respect to the connection $\Gamma^{\alpha}_{\phantom{\alpha}\mu\nu}$, the principle of last action gives~\cite{JHK1}
\begin{equation*}
	\delta_{\Gamma}S_{f(Q)}=0\ .
	\end{equation*}
The connection field equations,   with  the variation of  connection vanishing on domain boundary,   are
\begin{equation}\label{26}
\boxed{
\nabla_{\mu}\nabla_{\nu}\bigl(\sqrt{-g}f_{Q}P^{\mu\nu}_{\phantom{\mu\nu}\alpha}\bigr)=0
}\ .
\end{equation}
Constraints on the symmetric teleparallel theory are obtained by the vanishing of  variations with respect to  the Lagrange multipliers, i.e.,
\begin{equation}
\delta_{\lambda_{\alpha}^{\phantom{\alpha}\beta\mu\nu}}S_{f(Q)}=0\Rightarrow R^{\alpha}_{\phantom{\alpha}\beta\mu\nu}=0\ ,
\end{equation}
and 
\begin{equation}
\delta_{\lambda_{\alpha}^{\phantom{\alpha}\mu\nu}}S_{f(Q)}=0\Rightarrow T^{\alpha}_{\phantom{\alpha}\mu\nu}=0\ .
\end{equation}
The two sets of above field equations give the $f(Q)$ gravity dynamics.

\section{Linearized $f(Q)$ gravity in any gauge}\label{C}
In order to  linearize the non-metric gravity described by  action~\eqref{13}, let us  perturb around the Minkowski spacetime both the metric  and connection up to first order. These variables are  considered independent. It is   
\begin{equation}\label{40}
g_{\mu\nu}=\eta_{\mu\nu}+h_{\mu\nu}\quad\text{and}\quad\Gamma^{\alpha}_{\phantom{\alpha}\mu\nu}={\Gamma^{\alpha}}^{(0)}_{\mu\nu}+{\Gamma^{\alpha}}^{(1)}_{\mu\nu}\ ,
\end{equation}
where $\vert h_{\mu\nu}\vert \ll 1$ and $\vert{\Gamma^{\alpha}}^{(1)}_{\mu\nu}\vert\ll 1$,  with
\begin{equation}\label{41}
{\Gamma^{\alpha}}^{(0)}_{\mu\nu}=0\ ,
\end{equation}
because the non-metric connection disappears to zero-order when gravity is absent, having to reproduce the flat spacetime. For any gauge, to the first order of metric and connection perturbations, we get the following linear corrections 
\begin{equation}\label{42}
Q_{\alpha\mu\nu}^{(1)}=\partial_{\alpha}h_{\mu\nu}-2\,\Gamma^{(1)}_{(\mu|\alpha|\nu)}\ ,
\end{equation}
\begin{equation}\label{43}
{Q^{\alpha}}^{(1)}=\partial^{\alpha}h-2\,{\Gamma^{\lambda\phantom{\lambda}\alpha}_{\phantom{\lambda}\lambda}}^{(1)}\ ,
\end{equation}
\begin{equation}\label{44}
{\widetilde{Q}}^{\alpha(1)}=\partial_{\beta}h^{\alpha\beta}-2\,{\Gamma^{(\beta\phantom{\beta}\alpha)}_{\phantom{\beta}\beta}}^{(1)}\ ,
\end{equation}
\begin{equation}\label{45}
{B^{\alpha}}^{(1)}={Q^{\alpha}}^{(1)}-{\widetilde{Q}}^{\alpha(1)}=\partial^{\alpha}h-\partial_{\beta}h^{\alpha\beta}+{\Gamma^{\alpha\phantom{\beta}\beta}_{\phantom{\alpha}\beta}}^{(1)}-{\Gamma^{\beta\phantom{\beta}\alpha}_{\phantom{\beta}\beta}}^{(1)}\ ,
\end{equation}
\begin{equation}\label{46}
{L^{\alpha}_{\phantom{\alpha}\mu\nu}}^{(1)}=\frac{1}{2}\partial^{\alpha}h_{\mu\nu}-\partial_{(\mu}h^{\alpha}_{\phantom{\alpha}\nu)}+{\Gamma^{\alpha}_{\phantom{\alpha}\mu\nu}}^{(1)}\ ,
\end{equation}
\begin{multline}\label{47}
{P^{\alpha}_{\phantom{\alpha}\mu\nu}}^{(1)}=-\frac{1}{4}\partial^{\alpha}h_{\mu\nu}+\frac{1}{2}\partial_{(\mu}h^{\alpha}_{\phantom{\alpha}\nu)}+\frac{1}{4}\bigl(\partial^{\alpha}h-\partial_{\beta}h^{\alpha\beta}\bigr)\eta_{\mu\nu}-\frac{1}{4}\delta^{\alpha}_{(\mu}\partial_{\nu)}h\\
+\frac{1}{4}\bigl({\Gamma^{\alpha\phantom{\beta}\beta}_{\phantom{\alpha}\beta}}^{(1)}-{\Gamma^{\beta\phantom{\beta}\alpha}_{\phantom{\beta}\beta}}^{(1)}\bigr)\eta_{\mu\nu}-\frac{1}{2}{\Gamma^{\alpha}_{\phantom{\alpha}\mu\nu}}^{(1)}+\frac{1}{2}\delta^{\alpha}_{(\mu}{\Gamma^{\lambda}_{\phantom{\lambda}\lambda|\nu)}}^{(1)}\ .
\end{multline}
Expanding $f(Q)$ as 
\begin{equation}\label{48}
f(Q)=f(0)+f_{Q}(0)Q+O(Q^2)\ ,
\end{equation} 
we get
\begin{equation}\label{49}
f(Q)^{(0)}=f(0)=0\quad\text{and}\quad f(Q)^{(1)}=f_{Q}(0)Q^{(1)}=0\ ,
\end{equation}
\begin{equation}\label{49_1}
f_{Q}^{(0)}=f_{Q}(0)\quad\text{and}\quad f_{Q}^{(1)}=0\ ,
\end{equation}
assuming $f(0)=0$, and $f(Q)^{(1)}=0$ because, in $Q$, survive only second order terms in $h_{\mu\nu}$ and ${\Gamma^{\alpha}}^{(1)}_{\mu\nu}$.
Field Eqs.~\eqref{22} in vacuum, up to first order in $h$, become
\begin{equation}\label{50}
\partial_{\alpha}{P^{\alpha\mu}_{\phantom{\alpha\mu}\nu}}^{(1)}=0\ ,
\end{equation}
while  connection equations of motion~\eqref{26} at first order yield
\begin{equation}\label{51}
\partial_{\mu}\partial_{\alpha}{P^{\alpha\mu}_{\phantom{\alpha\mu}\nu}}^{(1)}=0\ ,
\end{equation}
which does not add any other constraints either on the metric or on the connection with respect to Eq.~\eqref{50}.  Finally, linearized field Eqs.~\eqref{50} in vacuum, in {\em any  gauge}, become
\begin{equation}\label{52}
\Box h_{\mu\nu}-2\,\partial^{\alpha}\partial_{(\mu}h_{\alpha|\nu)}+\partial_{\mu}\partial_{\nu}h+2\,\partial_{\alpha}{\Gamma^{\alpha}_{\phantom{\alpha}\mu\nu}}^{(1)}-2\,\partial_{(\mu}{\Gamma^{\lambda}_{\phantom{\lambda}\lambda|\nu)}}^{(1)}=0\ .
\end{equation}
Taking into account the absence of  torsion for the connection $\Gamma$,  it becomes symmetric,  namely
\begin{equation}\label{53}
T^{\alpha}_{\phantom{\alpha}\mu\nu}\bigl[\Gamma\bigr]=0\Rightarrow\Gamma^{\alpha}_{\phantom{\alpha}[\mu\nu]}=0\ .
\end{equation}
This result has been already used in the previous equations. Since we are in STG, the flatness of  connection gives  additional constraints, that is 
\begin{equation}\label{54}
R^{\alpha}_{\phantom{\alpha}\beta\mu\nu}\bigl[\Gamma\bigr]=0 \Rightarrow \Gamma^{\alpha}_{\phantom{\alpha}\beta\nu,\mu}-\Gamma^{\alpha}_{\phantom{\alpha}\beta\mu,\nu}+\Gamma^{\alpha}_{\phantom{\alpha}\lambda\mu}\Gamma^{\lambda}_{\phantom{\lambda}\beta\nu}-\Gamma^{\alpha}_{\phantom{\alpha}\lambda\nu}\Gamma^{\lambda}_{\phantom{\lambda}\beta\mu}=0\ .
\end{equation}
To the first order,  constraints~\eqref{54} become
\begin{equation}\label{55}
{\Gamma^{\alpha}_{\phantom{\alpha}\beta\nu,\mu}}^{(1)}={\Gamma^{\alpha}_{\phantom{\alpha}\beta\mu,\nu}}^{(1)}\ ,
\end{equation}
and by contracting the $\alpha$ and $\mu$ indices gives 
\begin{equation}\label{56}
\partial_{\alpha}{\Gamma^{\alpha}_{\phantom{\alpha}\beta\nu}}^{(1)}=\partial_{\nu}{\Gamma^{\alpha}_{\phantom{\alpha}\beta\alpha}}^{(1)}\ ,
\end{equation}
that, from the symmetry of  connection,  implies 
\begin{equation}\label{57}
2\partial_{\alpha}{\Gamma^{\alpha}_{\phantom{\alpha}\beta\nu}}^{(1)}=2\partial_{(\beta}{\Gamma^{\alpha}_{\phantom{\alpha}\alpha|\nu)}}^{(1)}\ .
\end{equation}
Finally, from~\eqref{57},  Eqs.~\eqref{52} are further simplified taking the following form~\cite{HPUS, SFSGS} 
\begin{equation}\label{58}
\boxed{
\Box h_{\mu\nu}-2\,\partial^{\alpha}\partial_{(\mu}h_{\alpha|\nu)}+\partial_{\mu}\partial_{\nu}h=0
}\ ,
\end{equation}%\label{59}
which are exactly the same differential equations  obtained by linearizing the $f(Q)$ field equations in the coincident gauge ~\cite{SCN2}.  This means that, in {\em any gauge}, to the first order  in  perturbations of  metric  and  connection, the linearized field equations in vacuum are independent of the connection adopted, because the linearized connection terms in Eq.~\eqref{52} vanish.  Furthermore, Eqs.~\eqref{58} are gauge-invariant, because under gauge transformation $x^{\prime\alpha}=x^{\alpha}+\epsilon^{\alpha}(x)$ with infinitesimal vector $\epsilon^{\alpha}$, remain unchanged. Indeed, according to the transformed perturbation  
\begin{equation}\label{59_1}
h^{\prime}_{\mu\nu}(x^{\prime})=h_{\mu\nu}(x)-2\partial_{(\mu}\epsilon_{\nu)}(x)\ ,
\end{equation}
 Eqs.~\eqref{58} transform as 
\begin{multline}\label{59_2}
\Box h^{\prime}_{\mu\nu}-2\,\partial^{\alpha}\partial_{(\mu}h^{\prime}_{\alpha|\nu)}+\partial_{\mu}\partial_{\nu}h^{\prime}=\Box h_{\mu\nu}-2\,\partial^{\alpha}\partial_{(\mu}h_{\alpha|\nu)}+\partial_{\mu}\partial_{\nu}h\\
+2\Big[\eta^{\alpha\beta}\partial_{\alpha}\partial_{\beta}\partial_{(\mu}\epsilon_{\nu)}+\partial^{\alpha}\partial_{\mu}\partial_{(\alpha}\epsilon_{\nu)}+\partial^{\alpha}\partial_{\nu}\partial_{(\alpha}\epsilon_{\mu)}-\eta^{\alpha\beta}\partial_{\mu}\partial_{\nu}\partial_{(\beta}\epsilon_{\alpha)}\Bigr]\ .
\end{multline}
The last line of  Eqs.~\eqref{59_2} cancels out and finally we obtain the gauge invariance  of linear field equations, namely
\begin{equation}\label{59_3}
\Box h^{\prime}_{\mu\nu}-2\,\partial^{\alpha}\partial_{(\mu}h^{\prime}_{\alpha|\nu)}+\partial_{\mu}\partial_{\nu}h^{\prime}=\Box h_{\mu\nu}-2\,\partial^{\alpha}\partial_{(\mu}h_{\alpha|\nu)}+\partial_{\mu}\partial_{\nu}h\ .
\end{equation}

\section{Gravitational wave solutions}\label{D} 
Let  us now use the spatial Fourier transformation, in order to search for wave-like solutions of Eqs.~\eqref{58}.  We seek for solutions as %the field Eqs.~\eqref{30} in the Fourier $\bm{k}$-vector space, according to the following waves expansion 
\begin{equation}\label{64_5}
h_{\mu\nu}(x)=\frac{1}{(2\pi)^{3/2}}\int\,d^{3}k\;\big(\widetilde{h}_{\mu\nu}\bigl(\bm{k}\bigr)e^{ik\cdot x}+ c.c.\bigr)\ ,
\end{equation}
with $\bm{k}$  the spatial wave vector and $c.c.$ the complex conjugate. Afterward, the field  Eqs.~\eqref{58}, in the momentum space, become
\begin{equation}\label{31}
\tilde{F}_{\mu\nu}=k^{2}\tilde{h}_{\mu\nu}-k_{\mu}k^{\alpha}\tilde{h}_{\alpha\nu}-k_{\nu}k^{\alpha}\tilde{h}_{\alpha\mu}+k_{\mu}k_{\nu}\tilde{h}=0\ ,
\end{equation}
where c.c. terms have been omitted.
Its trace  reads as
\begin{equation}\label{32}
k^{2}\tilde{h}-k^{\alpha}k^{\beta}\tilde{h}_{\alpha\beta}=0\ .
\end{equation}
Let us now consider a wave propagating in the $+z$ direction with the wave vector  $k^{\mu}=\bigl(\omega,0,0,k_{z}\bigr)$, where $k^{2}=\omega^{2}-k_{z}^{2}$. Thus the wave expansion~\eqref{64_5} becomes
\begin{equation}\label{43}
h_{\mu\nu}(z,t)=\frac{1}{\sqrt{2\pi}}\int dk_{z}\Bigl(\tilde{h}_{\mu\nu}(k_{z})e^{i(\omega t-k_{z}z)}+c.c.\Bigr)\ .
\end{equation}
 The ten components of linear field Eqs.~\eqref{31}, in the $k$-space, take the following form 
\begin{equation}\label{33}
\begin{aligned}
\tilde{F}_{00}&=\bigl(\omega^{2}+k_{z}^{2}\bigr)\tilde{h}_{00}+2\omega k_{z}\tilde{h}_{03}-\omega^{2}\tilde{h}=0\ ,\\
\tilde{F}_{01}&=k_{z}^{2}\tilde{h}_{01}+\omega k_{z}\tilde{h}_{13}=0\ ,\\
\tilde{F}_{02}&=k_{z}^{2}\tilde{h}_{02}+\omega k_{z}\tilde{h}_{23}=0\ ,\\
\tilde{F}_{03}&=\omega k_{z}\bigl(\tilde{h}_{00}-\tilde{h}_{33}-\tilde{h}\bigr)=0\ ,\\
\tilde{F}_{11}&=k^{2}\tilde{h}_{11}=0\ ,\\
\tilde{F}_{12}&=k^{2}\tilde{h}_{12}=0\ ,\\
\tilde{F}_{13}&=\omega k_{z}\tilde{h}_{01}+\omega^{2}\tilde{h}_{13}=0\ ,\\
\tilde{F}_{22}&=k^{2}\tilde{h}_{22}=0\ ,\\
\tilde{F}_{23}&=\omega k_{z}\tilde{h}_{02}+\omega^{2}\tilde{h}_{23}=0\ ,\\
\tilde{F}_{33}&=2\omega k_{z}\tilde{h}_{03}+\bigl(\omega^{2}+k_{z}^{2}\bigr)\tilde{h}_{33}+k_{z}^{2}\tilde{h}=0\,.
\end{aligned}
\end{equation}
The trace~\eqref{32}  becomes
\begin{equation}\label{34}
\omega^{2}\tilde{h}_{00}+2\omega k_{z}\tilde{h}_{03}+k_{z}^{2}\tilde{h}_{33}-\bigl(\omega^{2}-k_{z}^{2}\bigr)\tilde{h}=0\ ,
\end{equation}
where $\tilde{h}$ is the trace of the metric perturbation $h_{\mu\nu}$ in the momentum space given by
\begin{equation}
\tilde{h}=\tilde{h}_{00}-\tilde{h}_{11}-\tilde{h}_{22}-\tilde{h}_{33}\ .
\end{equation}
We first solve the set of Eqs.~\eqref{33} and~\eqref{34} for $k^{2}\neq 0$. It is straightforward to derive the following solution with four independent variables $\tilde{h}_{01}$, $\tilde{h}_{02}$, $\tilde{h}_{03}$ and $\tilde{h}_{00}$, namely 
\begin{equation}\label{35}
\begin{aligned}
\tilde{h}_{11}&=\tilde{h}_{12}=\tilde{h}_{22}=0\ ,\\
\tilde{h}_{13}&=-\frac{k_{z}}{\omega}\tilde{h}_{01}\ ,\\
\tilde{h}_{23}&=-\frac{k_{z}}{\omega}\tilde{h}_{02}\ ,\\
\tilde{h}_{33}&=-2\frac{k_{z}}{\omega}\tilde{h}_{03}-\frac{k_{z}^{2}}{\omega^{2}}\tilde{h}_{00}\ .
\end{aligned}
\end{equation}
Therefore, for $k^{2}\neq 0$, we obtain four degrees of freedom,  whose physical nature will be studied below via the geodesic deviation.

Then, for waves with $k^{2}=0$, the solution of Eqs.~\eqref{33} and \eqref{34}, where $\omega=k_{z}$, depends on six independent variables $\tilde{h}_{12}$, $\tilde{h}_{11}$, $\tilde{h}_{01}$, $\tilde{h}_{02}$, $\tilde{h}_{03}$ and $\tilde{h}_{00}$, namely 
\begin{equation}\label{36}
\begin{aligned}
\tilde{h}_{22}&=-\tilde{h}_{11}\ ,\\
%\tilde{h}_{12}&=\text{indeterminate}\ ,\\
\tilde{h}_{13}&=-\tilde{h}_{01}\ ,\\
\tilde{h}_{23}&=-\tilde{h}_{02}\ ,\\
\tilde{h}_{33}&=-2\tilde{h}_{03}-\tilde{h}_{00}\ ,
\end{aligned}
\end{equation}
and hence, six degrees of freedom which will be analyzed  again through the geodesic deviation.  Even if $f(Q)$ gravity seems to have from four to six degrees of freedom, we it is possible to show  that only two propagate. 

\section{Geodesic deviation in $f(Q)$ gravity}\label{E}
Let $x^{\alpha}(\tau)$ be a timelike curve on a manifold whose timelike tangent vector is $u^{\alpha}=dx^{\alpha}/d\tau$ with positive norm $\Vert u\Vert>0$ and $\tau$ the proper time.  In $f(Q)$ non-metric gravity, the equation of motion for a free test particle is 
\begin{equation}\label{60}
\boxed{
\frac{D^{\prime}u^{\lambda}}{d\tau}=L^{\lambda}_{\phantom{\lambda}\alpha\beta}\frac{dx^{\alpha}}{d\tau}\frac{dx^{\beta}}{d\tau}
}\ ,
\end{equation}
where 
\begin{equation}\label{61}
\frac{D^{\prime}}{d\tau}=u^{\alpha}\nabla_{\alpha}\ ,
\end{equation}
which is a force equation for the presence of the extra-force term on the right-hand side, see Appendix~\ref{H1} for details. Equivalently  Eq.~\eqref{60} can be written as the geodesic equation
\begin{equation}\label{61_1}
\frac{du^{\lambda}}{d\tau}+\genfrac\{\}{0pt}{1}{\lambda}{\alpha\beta}u^{\alpha}u^{\beta}=0\ ,
\end{equation}
that is,  free point particles follow the timelike metric geodesic but not the autoparallel curves with respect to the connection. As a consequence of the zero Levi-Civita (LC) covariant divergence on the left-hand side of  field Eqs~\eqref{22},  on-shell, we have the LC covariant conservation of  $T^{\mu}_{\phantom{\mu}\nu}$ on-shell. The geodesic deviation equation in $f(Q)$ gravity is 
\begin{equation}\label{62}
\boxed{
\frac{D^{\prime 2}\eta^{\mu}}{d\tau^{2}}-2u^{\alpha}L^{\mu}_{\phantom{\mu}\alpha\beta}\frac{D^{\prime}\eta^{\beta}}{d\tau}=\nabla_{\nu}L^{\mu}_{\phantom{\mu}\alpha\beta}\eta^{\nu}u^{\alpha}u^{\beta}
}\ ,
\end{equation}
where $\eta^{\mu}$ is the infinitesimal displacement vector connecting two nearby world lines. It is the  {\it deviation} or {\it separation vector}, see Appendix~\ref{H2} for details.  Now we rewrite the geodesic deviation  Eq.~\eqref{62} in a local Lorentz frame (LLF) that is in a coordinate system $\{x^{\hat{\alpha}}\}$ such as, in a point $\mathcal{P}_{0}$, the origin of the frame, the metric tensor $g_{\mu\nu}$ reads~\cite{CMW, MAGGIORE, MTW, CARROL} 
\begin{equation}
ds^2=dx^{\hat{0}^2}-\delta_{\hat{i}\hat{j}}dx^{\hat{i}}dx^{\hat{j}}+O(\vert x^{\hat{j}}\vert^2)dx^{\hat{\alpha}}dx^{\hat{\beta}}\ ,
\end{equation}
likewise such that 
\begin{equation}\label{62_01}
g_{\hat{\mu}\hat{\nu}}(\mathcal{P}_{0})=\eta_{\hat{\mu}\hat{\nu}}\quad\text{and}\quad\partial_{\hat{\alpha}}g_{\hat{\mu}\hat{\nu}}(\mathcal{P}_{0})=0\ ,
\end{equation}
 implying, in symmetric teleparallel theories, 
\begin{equation}\label{62_02}
\Gamma^{\hat{\alpha}}_{\phantom{\alpha}\hat{\mu}\hat{\nu}}(\mathcal{P}_{0})=L^{\hat{\alpha}}_{\phantom{\alpha}\hat{\mu}\hat{\nu}}(\mathcal{P}_{0})=0\ ,
\end{equation}
from the second relation of Eq.~\eqref{40}.
In particular, let us choose a proper reference frame of a freely falling test particle $A$, located at the  origin $\mathcal{P}_{0}$, and  another neighboring freely falling test particle $B$. The particle $B$ is  initially at rest with respect to $A$ and  can be described by the same proper frame to lowest order. Then, let us consider the 4-vector $\eta^{\hat{\alpha}}=(0,\vec{\chi)}$, where the spatial displacement $\chi^{\hat{i}}$ represents the spatial coordinate of the particle $B$ relative to particle $A$,  located at origin. The comoving 4-velocity of $A$ is $u_{A}^{\hat{\alpha}}=(1,0,0,0)$, while to the first order in $h$, the comoving 4-velocity  of $B$ is $u_{B}^{\hat{\alpha}}=(1,0,0,0)$.  To the first order in the  perturbation $h_{\mu\nu}$,  we can identify the proper time with the coordinate time, that is 
\begin{equation}\label{62_03}
t=\tau+O(h)\ .
\end{equation}
In our freely falling frame, the following identity is fulfilled,
\begin{equation}\label{62_04}
\partial_{\hat{0}}\Gamma^{\hat{i}}_{\phantom{i}\hat{0}\hat{j}}(\mathcal{P}_{0})=\partial_{\hat{0}}L^{\hat{i}}_{\phantom{i}\hat{0}\hat{j}}(\mathcal{P}_{0})=0\ ,
\end{equation}
and the expansion 
\begin{equation}\label{62_05}
\frac{D^{\prime 2}\chi^{\hat{i}}}{d\tau^2}=\frac{d^{2}\chi^{\hat{i}}}{dt^2}%+\partial_{\hat{0}}L^{\hat{i}}_{\phantom{i}\hat{0}\hat{j}}\chi^{\hat{j}}
+O(h^2)\ ,
\end{equation}
has to hold.
Thus, from Eqs.~\eqref{62}--\eqref{62_05}, the geodesic deviation equation~\eqref{62} in the proper reference frame reduces to 
\begin{equation}\label{64}
\boxed{
\ddot{\chi}^{\hat{i}}=2{\partial_{[\hat{j}}L^{\hat{i}}_{\phantom{\mu}\hat{0}|\hat{0}]}}^{(1)}{\chi^{\hat{j}}}
}\ ,
\end{equation} 
where the dot stands for the derivative with respect to the coordinate time $t$.  In any gauge, from the linearized disformation tensor~\eqref{46}, we obtain 
\begin{equation}\label{63}
{\partial_{\nu}L^{\mu}_{\phantom{\mu}\alpha\beta}}^{(1)}=\frac{1}{2}\Bigl(\partial_{\nu}\partial^{\mu}h_{\alpha\beta}-2\partial_{\nu}\partial_{(\alpha}h^{\mu}_{\phantom{\mu}\beta)}+2\partial_{\nu}{\Gamma^{\mu}_{\phantom{\mu}\alpha\beta}}^{(1)}\Bigr)\ ,
\end{equation}
and, from Eq.~\eqref{55}, we find 
\begin{equation}
2{\partial_{[\nu}L^{\mu}_{\phantom{\mu}\alpha|\beta]}}^{(1)}=\frac{1}{2}\Bigl(\partial_{\alpha}\partial_{\beta}h^{\mu}_{\phantom{\mu}\nu}+\partial_{\nu}\partial^{\mu}h_{\alpha\beta}-\partial^{\mu}\partial_{\beta}h_{\alpha\nu}-\partial_{\nu}\partial_{\alpha}h^{\mu}_{\phantom{\mu}\beta}\Bigr)\ ,
\end{equation}
that is, the contribution of  linearized connection disappears.  For our components in any gauge, we get 
\begin{equation}\label{65}
2{\partial_{[j}L^{i}_{\phantom{\mu}0|0]}}^{(1)}=\frac{1}{2}\Bigl(\partial_{0}\partial_{0}h^{i}_{\phantom{\mu}j}+\partial_{j}\partial^{i}h_{00}-\partial^{i}\partial_{0}h_{0j}-\partial_{j}\partial_{0}h^{i}_{\phantom{\mu}0}\Bigr)\ .
\end{equation}
From the gauge invariance of  Eq.~\eqref{65},  we have 
\begin{equation}
2{\partial_{[j}L^{i}_{\phantom{\mu}0|0]}}^{(1)}=2{\partial_{[\hat{j}}L^{\hat{i}}_{\phantom{\mu}\hat{0}|\hat{0}]}}^{(1)}\ ,
\end{equation}
which allows us to put Eq.~\eqref{64} into the following form
\begin{equation}\label{66_0}
\boxed{
\ddot{\chi}^{\hat{i}}=2{\partial_{[j}L^{i}_{\phantom{\mu}0|0]}}^{(1)}{\chi^{\hat{j}}}
}\ ,
\end{equation} 
which is the geodesic deviation equation of $fQ)$ gravity, in the proper reference frame of freely falling particles where ${\partial_{[j}L^{i}_{\phantom{\mu}0|0]}}^{(1)}$ is expressed in {\em any gauge},  as  the metric perturbations $h_{\mu\nu}$ from Eq.~\eqref{65}.  Eq.~\eqref{66_0} can be regarded as the relative acceleration between the two freely falling point particles. 

\section{Polarization of Gravitational Waves}\label{F}
Let us now consider the following expansion to the first order of  spatial separation vector $\chi^{\hat{i}}=(\chi_{x},\chi_{y},\chi_{z})$ in the LLF
\begin{equation}\label{66_00}
\chi^{\hat{i}}(t)=\chi^{(0)\hat{i}}+\delta x^{\hat{i}}(t)\ ,
\end{equation}
where $\chi^{(0)\hat{i}}=(x_{0},y_{0},z_{0})$ is the initial displacement, i.e., the initial relative position of the particle $B$ with respect to the particle $A$  and $\delta x^{\hat{i}}=(\delta x, \delta y, \delta y)$ the linear perturbation of the spatial deviation vector. The relative variation of the linear perturbation $\delta x^{\hat{i}}$ with respect to the displacement $\chi^{(0)\hat{i}}$ is of the same order of $h$, namely
\begin{equation}
\frac{\vert \delta x^{\hat{i}}\vert}{\vert \chi^{(0)\hat{i}}\vert}=O(h)\ .
\end{equation}
Inserting Eq.~\eqref{65} into Eq.~\eqref{64}, the linear system of differential equations, for a wave traveling along positive $z$-axis in a proper local reference frame, to the  first order according to Eq.~\eqref{66_00} reads as
\begin{equation}\label{66}
\left\{
\begin{array}{lr}
\ddot{\delta x}=-\frac{1}{2}\bigl(h_{11,00}\bigr)x_{0}-\frac{1}{2}\bigl(h_{12,00}\bigr)y_{0}+\frac{1}{2}\bigl(h_{01,03}-h_{13,00}\bigr)z_{0}\\
\ddot{\delta y}=-\frac{1}{2}\bigl(h_{12,00}\bigr)x_{0}-\frac{1}{2}\bigl(h_{22,00}\bigr)y_{0}+\frac{1}{2}\bigl(h_{02,03}-h_{23,00}\bigr)z_{0}\\
\ddot{\delta z}=\frac{1}{2}\bigl(h_{01,03}-h_{13,00}\bigr)x_{0}+\frac{1}{2}\bigl(h_{02,03}-h_{23,00}\bigr)y_{0}+\frac{1}{2}\bigl(2h_{03,03}-h_{33,00}-h_{00,33}\bigr)z_{0}
\end{array}\,
\right.\ .
\end{equation}
We have taken  into account that 
\begin{equation}
{\partial_{j}L^{i}_{\phantom{i}00}}^{(1)}=-{\partial_{j}L_{i00}}^{(1)}\ ,
\end{equation}
where  perturbations $h_{\mu\nu}$ are expressed in an arbitrary gauge.
For a plane wave at fixed $k_{z}$, with the  angular frequency $\omega$, related by $k^{2}=\omega^{2}-k_{z}^{2}$,  as 
\begin{equation}
h_{\mu\nu}=\widetilde{h}_{\mu\nu}\bigl(k_{z}\bigr)e^{i(\omega t-k_{z}x)}+c.c.\ ,
\end{equation}
system~\eqref{66} yields
\begin{equation}\label{66_11}
\left\{
\begin{array}{ll}
\ddot{\delta x}(t)=\Bigl[\frac{1}{2}\omega^{2}\tilde{h}_{11}x_{0}+\frac{1}{2}\omega^{2}\tilde{h}_{12}y_{0}+\frac{1}{2}\bigl(\omega k_{z}\tilde{h}_{01}+\omega^2\tilde{h}_{13}\bigr)z_{0}\Bigr]e^{i(\omega t-k_{z}x)}\\
\ddot{\delta y}(t)=\Bigl[\frac{1}{2}\omega^2\tilde{h}_{12}x_{0}+\frac{1}{2}\omega^2\tilde{h}_{22}y_{0}+\frac{1}{2}\bigl(\omega k_{z}\tilde{h}_{02}+\omega^2\tilde{h}_{23}\bigr)z_{0}\Bigr]e^{i(\omega t-k_{z}x)}\\
\ddot{\delta z}(t)=\Bigl[\frac{1}{2}\bigl(\omega k_{z}\tilde{h}_{01}+\omega^2\tilde{h}_{13}\bigr)x_{0}+\frac{1}{2}\bigl(\omega k_{z}\tilde{h}_{02}+\omega^2\tilde{h}_{23}\bigr)y_{0}\\
\quad\quad\quad\quad\quad\quad\quad\quad\quad\quad+\frac{1}{2}\bigl(2\omega k_{z}\tilde{h}_{03}+\omega^2\tilde{h}_{33}+k_{z}^{2}\tilde{h}_{00}\bigr)z_{0}\Bigr]e^{i(\omega t-k_{z}x)}
\end{array}
\right. +c.c.\ .
\end{equation}
In the case $k^{2}=M^{2}\neq 0$, we insert  solution~\eqref{35} in  system~\eqref{66_11}, assuming fixed $k_{z}$. It takes the form
\begin{equation}\label{67}
\left\{
\begin{array}{lr}
\ddot{\delta x}=0\\
\ddot{\delta y}=0\\
\ddot{\delta z}=0
\end{array}\ .
\right.
\end{equation}
Also, imposing the initial conditions, such as the initial displacement perturbation $\delta x^{\hat{i}}(0)=0$ and its initial velocity $\dot{\delta x}^{\hat{i}}(0)=0$, after a double integration with respect to $t$ of  system~\eqref{67},  we obtain the solution
\begin{equation}\label{45}
\delta x(t)=0\ ,\quad \delta y(t)=0\ ,\quad \delta z(t)=0\ ,
\end{equation}
which is not a wave, i.e., there is no massive mode associated with $k^{2}\neq 0$. That is, when $k^{2}\neq 0$, none of the four degrees of freedom has a physical meaning.

On the other hand, in the case $k^{2}=0$,  implying $\omega=k_{z}$, by means of solution~\eqref{36}, the geodesic deviation equation, to the first order in perturbation for a fixed $k_{z}$ of a single plane wave~\eqref{66_11}, yields 
\begin{equation}\label{46_1}
\left\{
\begin{array}{lr}
\ddot{\delta x}=\frac{1}{2}\omega^{2}\bigl(\tilde{h}^{(+)}x_{0}+\tilde{h}^{(\times)}y_{0}\bigr)e^{i\omega(t-z)}\\
\ddot{\delta y}=\frac{1}{2}\omega^{2}\bigl(\tilde{h}^{(\times)}x_{0}-\tilde{h}^{(+)}y_{0}\bigr)e^{i\omega(t-z)}\\
\ddot{\delta z}=0
\end{array}
\right.+ c.c.\ ,
\end{equation}
where $\tilde{h}_{11}=\tilde{h}^{(+)}$ and $\tilde{h}_{12}=\tilde{h}^{(\times)}$. Hence, in any gauge, for $k^2=0$, only two degrees of freedom of the initial six survive exactly as in the case of GR. Then, after double integration with respect to $t$, the solution of system~\eqref{46_1} becomes
\begin{equation}\label{47}
\left\{
\begin{array}{lr}
\delta x(t)=-\frac{1}{2}\bigl(\tilde{h}^{(+)}x_{0}+\tilde{h}^{(\times)}y_{0}\bigr)e^{i\omega(t-z)}\\
\delta y(t)=-\frac{1}{2}\bigl(\tilde{h}^{(\times)}x_{0}-\tilde{h}^{(+)}y_{0}\bigr)e^{i\omega(t-z)}\\
\delta z(t)=0
\end{array}
\right.+ c.c.\ ,
\end{equation}
describing the response of a ring of freely falling masses hit by a gravitational wave. When we have the pure tensor mode $\tilde{h}^{(\times)}=0$, from Eq.~\eqref{47}, the solution is 
\begin{equation}
\left\{
\begin{array}{ll}
\delta x(t)=-\tilde{h}^{(+)}x_{0}\cos (\omega (t-z))\\
\delta y(t)=+\tilde{h}^{(+)}y_{0}\cos (\omega (t-z))
\end{array}
\right.\ ,
\end{equation} 
that is, the effect of gravitational wave is to distort the circle of particles into ellipses oscillating in a $+$ pattern. While in the pure tensor mode $\tilde{h}^{(+)}=0$, the ring distorts into ellipses oscillating in a $\times$ pattern rotated by $45$ degrees in a right-handed sense with respect to it, because the solution of  Eq.~\eqref{47} is 
\begin{equation}
\left\{
\begin{array}{ll}
\delta x(t)=-\tilde{h}^{(\times)}y_{0}\cos (\omega (t-z))\\
\delta y(t)=-\tilde{h}^{(\times)}x_{0}\cos (\omega (t-z))
\end{array}
\right.\ .
\end{equation} 
In summary, we have obtained the well-known plus and cross, massless, transverse, and spin 2 modes of GR. The metric perturbation induced by gravitational wave can be put into the form
\begin{equation}\label{48}
h_{\mu\nu}(z,t)=\frac{1}{\sqrt{2\pi}}\int d\omega\Bigl[\epsilon_{\mu\nu}^{(+)}\tilde{h}^{(+)}(\omega)+\epsilon_{\mu\nu}^{(\times)}\tilde{h}^{(\times)}(\omega)\Bigr]e^{i\omega (t-z)}+c.c.\ ,
\end{equation}
where $\epsilon_{\mu\nu}^{(+)}$ and $\epsilon_{\mu\nu}^{(\times)}$ are the polarization tensors defined as 
\begin{equation}\label{49}
\epsilon^{(+)}_{\mu\nu}=\frac{1}{\sqrt{2}}
\begin{pmatrix} 
0 & 0 & 0 & 0 \\
0 & 1 & 0 & 0 \\
0 & 0 & -1 & 0 \\
0 & 0 & 0 & 0
\end{pmatrix}\ ,
\end{equation}
\begin{equation}\label{49.1}
\epsilon^{(\times)}_{\mu\nu}=\frac{1}{\sqrt{2}}
\begin{pmatrix} 
0 & 0 & 0 & 0 \\
0 & 0 & 1 & 0 \\
0 & 1 & 0 & 0 \\
0 & 0 & 0 & 0
\end{pmatrix}\ ,
\end{equation}
which, under spatial rotations around the $z$-axes,  transform such that the helicity of perturbation field $h_{\mu\nu}$ is equal two. Hence without fixing any gauge in $f(Q)$ non-metric gravity, we obtain exactly the same GWs predicted by GR in the TT gauge. Finally a gravitational wave propagating along an arbitrary $\mathbf{k}$ direction in $f(Q)$ gravity, from Eq.~\eqref{48}, becomes
\begin{equation}\label{49.2}
\boxed{
h_{\mu\nu}(x)=\frac{1}{(2\pi)^{3/2}}\int d^3k\Bigl[\epsilon_{\mu\nu}^{(+)}\tilde{h}^{(+)}(\mathbf{k})+\epsilon_{\mu\nu}^{(\times)}\tilde{h}^{(\times)}(\mathbf{k})\Bigr]e^{i(\omega t-\mathbf{k}\cdot\mathbf{x})}+c.c.
}\ .
\end{equation}
Finally, considering  the total   six degrees of freedom,  only two with $k^2=0$ propagate, exactly $\tilde{h}^{(+)}$ and $\tilde{h}^{(\times)}$. The geodesic deviation equation is valid as long as the modulus of the spatial deviation vector $\vert \chi^{i}\vert$ is much smaller than typical scale over the gravitational field changes. Thus for a gravitational wave of wavelength $\lambda$ and a GWs detector of characteristic linear size $L$, such as resonant bar detector or ground-based (LIGO-VIRGO, Einstein Telescope) and space-based (LISA) interferometers, the following approximation 
\begin{equation}\label{49.3}
L\ll\lambda\ ,
\end{equation}
must hold~\cite{MAGGIORE}.  In conclusion, as in the case of $f(T)$ teleparallel gravity \cite{Bamba}, no further GW modes  than those of GR emerge.

\section{Discussion and Conclusions}\label{G}
In this paper,  we  considered the GW polarization modes  in $f(Q)$ non-metric gravity without fixing any  gauge. Through   the geodesic deviation equation, we analyzed   possible   massive and  massless scalar and tensor modes. The main result is that,  in any gauge,  the gravitational radiation of  $f(Q)$ gravity exhibits only tensor modes  without any scalar component. Specifically,  in the linearized  theory, only two  degrees of freedom propagate, that is  $h^{(+)}$ and $h^{(\times)}$. They uniquely encode the gravitational field. Furthermore, this radiation is transverse and massless, reproducing exactly the same plus and cross  polarizations of GR.  

We first perturbed the metric tensor and the affine connection around the Minkowski spacetime. We  thus derived  the field and connection equations of $f(Q)$ in vacuum at the first order. These equations    remain gauge invariant because the linear perturbation of  connection disappears. We therefore obtained, at the first order,  the same equations in vacuum obtained in the coincidence gauge.   This result is due to the fact that  even if the gauge is left free, the GW modes  remain unchanged.  Furthermore,  we verified that in our , symmetric, teleparallel, non-metric theory, the world lines of  free point particles follow  timelike geodesics and not  autoparallel curves as consequence of LC covariant conservation of the  energy-momentum tensor on-shell, namely $\mathcal{D}_{\alpha}T^{\alpha}_{\phantom{\alpha}\beta}=0$.  We hence obtained the particle equations of motion  using only the connection without imposing the LC connection. This is because we have equipped our spacetime manifold with a metric and  the symmetric teleparallel connection.  The equation for the deviations between the nearby geodesics was obtained in the same way using only our non-metric connection. Subsequently, we first solved the linearized field equations in vacuum in the momentum space by seeking for wave solutions.  After assuming a LLF, the comoving one with free particles in free fall, we solved the geodetic deviations equation thanks to the gauge invariance of the derivatives of  linearized disformation tensor $L^{\alpha(1)}_{\phantom{\alpha}\mu\nu}$,  which  allows us to express the metric perturbations $h_{\mu\nu}$ in any gauge. Finally, we showed that massive gravitational radiation modes with $k^2\neq 0$  do not propagate while massless radiation with $k^2=0$  propagate.  That is, when a massless GW hits a ring of particles, the first-order perturbations of their position from the center are transverse  tensors, exactly reproducing the two standard modes of  GR.  To the first order in $h$ and $\Gamma^{(1)}$, no further scalar mode oscillates either massive or massless,  {\em regardless of the gauge adopted}.

Experimentally, if, for example,  we take into account  as GW source, a  coalescence of BH-BH binary system, emitting  radiation with a frequency $\nu\sim 100\,\text{Hz}$ ($\lambda\sim 3000\,\text{km}$) at cosmological distances $R\sim 100\,\text{Mpc}$, the typical amplitude or strain is $h\sim\,10^{-21}$. The  arm length of the ground-based interferometers (LIGO-VIRGO) is $L\sim 3\text{--}4\,\text{km}$ so the approximation~\eqref{49.3}  holds and the strain falls within the sensitivity of  GW Earth observatories.  Additionally, the LISA space interferometer~\cite{LISA} will increase the sensitivity by a factor  10, $h\sim 10^{-22}$ allowing the detection of weaker GW signals in low frequencies between $10^{-4}$ and $10^{-1}$ Hz,  and the Einstein Telescope  will further improve the sensitivity allowing  to detect signals with even smaller strains $h\sim 10^{-23}$ in the frequency range $\nu=10^{0} \div 10^{3}\, \text{Hz}$~\cite{ET}.  However, neither GR nor $f(T)$ gravity~\cite{Aldrovandi,Bamba,Ong:2013qja, Izumi:2012qj} can be distinguished from the $f(Q)$ theory via GW measurements because they exhibit the same tensor modes. On the other hand,  $f(R)$ gravity  could exhibit detectable scalar radiation \cite{Caprep}.  Onnthe other hand, if we  add a boundary term to $f(Q)$, it is possible to prove that the theory exhibits an additional scalar mode like $f(R)$ \cite{Carmen}. The fact that the non-linear extensions $f(R)$, $f(T)$, and $f(Q)$ are not equivalent while their linear actions are equivalent, i.e., the geometric trinity of gravity, has consequences on the physical degrees of freedom and could have important consequences in cosmology in discriminating among concurring models~\cite{Koussour,Mandal,Naik,Vishwakarma,Shi,Ferreira,Bajardi,Koussour1}.  It is worth noticing that symmetric teleparallel theories could lead to an overall amplification of the stochastic spectrum, as recently pointed out  in  other modified theories of gravity~\cite{Odi1}. 

\appendix

\section{Appendices}\label{H}
\subsection{ The equations of motion for particles}\label{H1}
The matter energy-momentum tensor $T^{\alpha}_{\phantom{\alpha}\beta}$ for a perfect fluid, specifically for dust, in its rest frame is~\cite{RV, HKLOR, XLHL}   
\begin{equation}\label{100}
T^{\alpha}_{\phantom{\alpha}\beta}=\rho u^{\alpha}u_{\beta}\ ,
\end{equation}
where $\rho$ is the proper mass-energy density of a particle of mass $m$ located at $\vec{x}_{0}$ i.e.,
\begin{equation}
\rho(x)=\frac{m}{\sqrt{-g}}\delta^3\big[\vec{x}(\tau)-\vec{x}(0)\bigr]\ ,
\end{equation}
with $\delta^3$ the spatial Dirac delta, $\tau$ the proper time and $u^{\alpha}$ the comoving 4-velocity.   Performing the covariant derivative of Eq.~\eqref{100}, we have  
\begin{equation}\label{101}
\nabla_{\alpha}T^{\alpha}_{\phantom{\alpha}\beta}=\rho\bigl(\nabla_{\alpha}u^{\alpha}\bigr)u_{\beta}+\rho\,u^{\alpha}\nabla_{\alpha}u_{\beta}\ .
\end{equation}
We can write the field Eqs~\eqref{22} as
\begin{equation}\label{102}
F^{\alpha}_{\phantom{\alpha}\beta}=\kappa^{2}T^{\alpha}_{\phantom{\alpha}\beta}\ .
\end{equation}
where $F^{\alpha}_{\phantom{\alpha}\beta}$ is the left-hand side of Eqs.\eqref{22}.
We have the vanishing LC covariant divergence on-shell, i.e., for metric tensor $g_{\mu\nu}$ and connection $\Gamma^{\alpha}_{\phantom{\alpha}\mu\nu}$, solutions of Eqs.~\eqref{22} and~\eqref{26}, it is
\begin{equation}\label{103}
\mathcal{D}_{\alpha}F^{\alpha}_{\phantom{\alpha}\beta}=0\ ,
\end{equation}
where $\mathcal{D}$ is the covariant derivative with respect to the Christoffel symbols. This implies the LC covariant conservation of the energy-momentum tensor on-shell, namely
\begin{equation}
\mathcal{D}_{\alpha}T^{\alpha}_{\phantom{\alpha}\beta}=0\ .
\end{equation}
 Then, from the identity
\begin{equation}\label{103_1}
\nabla_{\alpha}F^{\alpha}_{\phantom{\alpha}\beta}=\mathcal{D}_{\alpha}F^{\alpha}_{\phantom{\alpha}\beta}-\frac{1}{2}Q_{\lambda}F^{\lambda}_{\phantom{\lambda}\beta}+L^{\lambda}_{\phantom{\lambda}\alpha\beta}F^{\alpha}_{\phantom{\alpha}\lambda}\ ,
\end{equation}
we derive the covariant divergence of stress-energy tensor 
\begin{equation}\label{104}
\nabla_{\alpha}T^{\alpha}_{\phantom{\alpha}\beta}=-\frac{1}{2}Q_{\lambda}T^{\lambda}_{\phantom{\lambda}\beta}+L^{\lambda}_{\phantom{\lambda}\alpha\beta}T^{\alpha}_{\phantom{\alpha}\lambda}\ .
\end{equation}
From Eq.~\eqref{101} and \eqref{104},  we have
\begin{equation}\label{105}
\rho\bigl(\nabla_{\alpha}u^{\alpha}\bigr)u_{\beta}+\rho\,g_{\sigma\beta}u^{\alpha}\nabla_{\alpha}u^{\sigma}+\rho Q_{\alpha\sigma\beta}u^{\alpha}u^{\sigma}=-\frac{1}{2}Q_{\lambda}T^{\lambda}_{\phantom{\lambda}\beta}+L^{\lambda}_{\phantom{\lambda}\alpha\beta}T^{\alpha}_{\phantom{\alpha}\lambda}\ .
\end{equation}
So, from the normalization condition of the four-velocity $u^{\alpha}u_{\alpha}=1$, differentiating it with respect to the non-metric connection, we get the following differential identity
\begin{equation}\label{106}
2u_{\alpha}\nabla_{\beta}u^{\alpha}+Q_{\beta\alpha\rho}u^{\alpha}u^{\rho}=0\ .
\end{equation}
Projecting Eqs.~\eqref{105} on the 3-space, orthogonal to the four-velocity by the projector operator $h^{\beta}_{\mu}=\delta^{\beta}_{\mu}-u^{\beta}u_{\mu}$ and from the differential relation~\eqref{106}, we obtain 
\begin{equation}\label{107}
g_{\sigma\mu}\frac{D^{\prime}u^{\sigma}}{d\tau}-\frac{1}{2}\bigl(Q_{\alpha\beta\sigma}+2L_{\sigma\beta\alpha}\bigr)u^{\alpha}u^{\beta}u^{\sigma}u_{\mu}+\bigl(L^{\sigma}_{\phantom{\sigma}\alpha\mu}+Q_{\alpha\mu}^{\phantom{\alpha\mu}\sigma}\bigr)u^{\alpha}u_{\sigma}=0\ .
\end{equation}
According to the following identities 
\begin{equation}\label{108}
\bigl(Q_{\alpha\beta\sigma}+2L_{\sigma\beta\alpha}\bigr)u^{\alpha}u^{\beta}u^{\sigma}=0\ ,
\end{equation}
and 
\begin{equation}\label{109}
L^{\sigma}_{\phantom{\sigma}\alpha\mu}+Q_{\alpha\mu}^{\phantom{\alpha\mu}\sigma}=-L_{\mu\alpha}^{\phantom{\mu\alpha}\sigma}\ ,
\end{equation}
Eqs.~\eqref{107} finally read as 
\begin{equation}\label{110}
\frac{D^{\prime}u^{\nu}}{d\tau}=L^{\nu}_{\phantom{\nu}\alpha\beta}u^{\alpha}u^{\beta}\ ,
\end{equation}
that is, from  Eq.~\eqref{61_1}, the equations of motion for  metric geodesic are governed  by Christoffel symbols and therefore by the metric tensor $g_{\mu\nu}$.

\subsection{ The geodesic deviation equation}\label{H2}
Let us calculate the second covariant derivative of the deviation vector connecting two nearby geodesics~\eqref{110},  i.e.,
\begin{equation}\label{200}
H^{\alpha}=\nabla_{u}\nabla_{u}\eta^{\alpha}\ ,
\end{equation}
where the covariant derivative along the tangent vector $u$ is 
\begin{equation}\label{201}
\nabla_{u}=u^{\alpha}\nabla_{\alpha}\ .
\end{equation}
In symmetric teleparallel theories, the $\eta=\eta^{\alpha}{\bf e}_{\alpha}$ and $u=u^{\beta}{\bf e}_{\beta}$ four-vectors commute in a coordinate bases $\{{\bf e}_{\alpha}\}$, since in a symmetric theory the torsion is zero, that is their commutator vanishes 
\begin{equation}\label{202}
[\eta,u]=0\ ,
\end{equation}
or equivalently 
\begin{equation}\label{203}
\eta^{\alpha}\nabla_{\alpha}u^{\beta}=u^{\alpha}\nabla_{\alpha}\eta^{\beta}\ ,
\end{equation}
or again 
\begin{equation}\label{204}
\nabla_{\eta}u^{\beta}=\nabla_{u}\eta^{\beta}\ .
\end{equation}
Now from~\eqref{200} and Eq.~\eqref{202} and the commutativity of the covariant derivatives $\nabla_{\rho}$ with respect to the connection that has neither torsion nor curvature, it is
\begin{equation}\label{205}
\begin{split}
H^{\alpha}&=\frac{{D^{\prime}}^{2}\eta^{\alpha}}{d\tau^2}=u^{\rho}\nabla_{\rho}\bigl(u^{\sigma}\nabla_{\sigma}\eta^{\alpha}\bigr)\\
&=\bigl(u^{\rho}\nabla_{\rho}\eta^{\sigma}\bigr)\big(\nabla_{\sigma}u^{\alpha}\bigr)+u^{\rho}\eta^{\sigma}\nabla_{\sigma}\nabla_{\rho}u^{\alpha}\\
&=\bigl(\eta^{\rho}\nabla_{\rho}u^{\sigma}\bigr)\big(\nabla_{\sigma}u^{\alpha}\bigr)+\eta^{\sigma}\nabla_{\sigma}\bigl(u^{\rho}\nabla_{\rho}u^{\alpha}\bigr)-\bigl(\eta^{\sigma}\nabla_{\sigma}u^{\rho}\bigr)\bigl(\nabla_{\rho}u^{\alpha}\bigr)\\
&=\eta^{\sigma}\nabla_{\sigma}\bigl(u^{\rho}\nabla_{\rho}u^{\alpha}\bigr)
\end{split}\ ,
\end{equation}
where in the third line we used the Leibniz rule and Eq.~\eqref{202} again.  Then, from the equation of motion~\eqref{110},  Eq.~\eqref{205} becomes
\begin{equation}\label{206}
\begin{split}
\frac{{D^{\prime}}^{2}\eta^{\nu}}{d\tau^2}&=\eta^{\sigma}\nabla_{\sigma}\bigl(L^{\nu}_{\phantom{\nu}\alpha\beta}u^{\alpha}u^{\beta}\bigr)\\
&=\nabla_{\sigma}L^{\nu}_{\phantom{\nu}\alpha\beta}\eta^{\sigma}u^{\alpha}u^{\beta}+2L^{\nu}_{\phantom{\nu}\alpha\beta}\bigl(\eta^{\sigma}\nabla_{\sigma}u^{\alpha}\bigr)u^{\beta}\\
&=\nabla_{\sigma}L^{\nu}_{\phantom{\nu}\alpha\beta}\eta^{\sigma}u^{\alpha}u^{\beta}+2L^{\nu}_{\phantom{\nu}\alpha\beta}\bigl(u^{\sigma}\nabla_{\sigma}\eta^{\alpha}\bigr)u^{\beta}
\end{split}\ ,
\end{equation} 
where, in the last line, we always used Eq.~\eqref{202}. Finally  Eq.~\eqref{206} can be put in the following form 
\begin{equation}
\frac{{D^{\prime}}^{2}\eta^{\nu}}{d\tau^2}=\nabla_{\sigma}L^{\nu}_{\phantom{\nu}\alpha\beta}\eta^{\sigma}u^{\alpha}u^{\beta}+2L^{\nu}_{\phantom{\nu}\alpha\beta}\frac{D^{\prime}\eta^{\alpha}}{d\tau}u^{\beta}\ ,
\end{equation}
that is the geodesic deviation equation of $f(Q)$ gravity. 

\section*{Acknowledgements}
Authors acknowledge the Istituto Nazionale di Fisica Nucleare (INFN) Sez.~di~Napoli, Iniziative Specifiche QGSKY and MOONLIGHT2, and the Istituto Nazionale di Alta Matematica (INdAM), gruppo GNFM, for the support.
This paper is based upon work from COST Action CA21136: {\it Addressing observational tensions in cosmology with systematics and fundamental physics} (CosmoVerse) supported by COST (European Cooperation in Science and Technology).

\end{document}